\documentclass[11pt]{article}
\usepackage{}
\usepackage{amsfonts}
\usepackage{amsmath,amssymb}
\usepackage{graphicx}
\usepackage{caption2}
\usepackage{epsfig}
\usepackage{subfigure}
\usepackage[dvips]{color}

\def\dref#1{(\ref{#1})}
\def\rm{\mathrm}
\begin{document}

\begin{center}
{\LARGE\bf{Distributed Robust Control of Linear Multi-Agent Systems
with Parameter Uncertainties} \footnote[1] {Zhongkui Li and Xiangdong Liu
are with the School of Automation, Beijing Institute of
Technology,\, Beijing 100081, China (e-mail:
zhongkli@gmail.com). Zhisheng Duan is with State Key Lab for Turbulence and Complex Systems,
Department of Mechanics and Aerospace Engineering, College of Engineering, Peking University,
Beijing 100871, China (e-mail:
duanzs@pku.edu.cn).
Lihua Xie is with the School of Electrical and
Electronic Engineering, Nanyang Technological University, Singapore,
639798 (e-mail: elhxieg@ntu.edu.sg).}}
\end{center}

\vskip 0.3cm \centerline{Zhongkui Li, Zhisheng Duan, Lihua Xie,
Xiangdong Liu}
\vskip 0.4cm
\centerline{\today}

\vskip 1cm

{\noindent \small {\bf  Abstract}: This paper considers the distributed robust
control problems of uncertain linear multi-agent systems with
undirected communication topologies. It is assumed that
the agents have identical nominal dynamics while subject to different
norm-bounded parameter uncertainties, leading to weakly heterogeneous multi-agent systems.
Distributed controllers are designed for both continuous- and discrete-time multi-agent systems,
based on the relative states of
neighboring agents and a subset of absolute states of the agents.
It is shown for both the continuous- and discrete-time cases that the distributed robust control
problems under
such controllers in the sense of quadratic stability are equivalent to the $H_\infty$ control problems
of a set of decoupled linear systems having the same dimensions as a single agent.
A two-step algorithm is presented to construct
the distributed controller for the continuous-time case, which does not involve any conservatism
and meanwhile decouples
the feedback gain design from the communication topology.
Furthermore, a sufficient existence condition
in terms of linear matrix inequalities is derived for
the distributed discrete-time controller.
Finally, the distributed robust $H_\infty$ control
problems of uncertain linear multi-agent systems subject to external disturbances
are discussed.

\vskip 0.2cm

}

\vskip 0.6cm
\section{Introduction}

The coordination control problems of multi-agent
systems have received increasing attention from various scientific
communities, for its broad applications
in such fields as satellite formation flying, sensor networks,
and air traffic control, to
name just a few \cite{olfati-saber2007consensus}.
Due to the spatial distribution of actuators, communication constraint,
and limited sensing capability of sensors,
centralized controllers are generally too expensive
or even infeasible to implement in practice.
Therefore, distributed control strategies based on only local information
have been proposed and extensively studied in the last decade.

Formation control of multiple autonomous vehicles is considered
in \cite{fax2004information}, where a Nyquist-like criterion is
derived. Distributed linear quadratic regulator (LQR) control
of a set of identical decoupled
dynamical systems is discussed in \cite{borrelli2008distributed}.
A decomposition approach is proposed in \cite{massioni2009distributed}
to solve the distributed $H_\infty$ control of identical
coupled linear systems. A general
framework of the consensus problem for networks of integrator agents
with fixed and switching topologies is addressed in
\cite{olfati-saber2004consensus}.
The conditions derived in \cite{olfati-saber2004consensus} are
further relaxed in \cite{ren2005consensus}.
The consensus problems of networks of double-
and high-order integrators are investigated in
\cite{ren2008consensus,xie2007consensus,ren2007high-order}.
Consensus algorithms are designed in \cite{li2010distributed,carli2009quantized} for a group
of agents with quantized communication links and limited data rate.
Distributed consensus control of multi-agent systems with general
linear dynamics is studied in
\cite{tuna2009conditions,scardovi2009synchronization,li2010consensus,ma2010necessary}.
Distributed $H_\infty$ consensus and control problems are
investigated in \cite{lin2008distributed,li2011hinf} for networks of
agents subject to external disturbances and model uncertainties.
Flocking algorithms are studied
in \cite{olfati2006flocking,su2009flocking}
for a group of autonomous agents.

The aforementioned works have a common assumption that the dynamics
of the agents are all identical, i.e., the multi-agent systems are homogeneous.
Such an assumption may be restrictive in many circumstances. It is necessary
and important to study the cooperative control problems of heterogenous multi-agent systems
consisting of nonidentical agents. Previous works along this line include
\cite{lestas2006scalable,lestas2007scalable,jonsson2010scalable,
wang2010distributed,das2010distributed}.
Specifically, scalar robust stability conditions are derived in
\cite{lestas2006scalable,lestas2007scalable} for a network of
heterogeneous agents by using the notion of S-hull. Similar results
are given in \cite{jonsson2010scalable}
using tools from the integral quadratic constraint (IQC).
In \cite{wang2010distributed}, a distributed controller based on the
internal model is designed for
the output regulation of heterogeneous linear
multi-agent systems. 
Neural adaptive tracking control
of first-order nonlinear systems with unknown dynamics and disturbances
is investigated in \cite{das2010distributed}.

This paper is concerned with the distributed robust control problems
of multi-agent systems with general linear dynamics subject to norm-bounded parameter
uncertainties. It is assumed that all the agents have the same nominal dynamics but subject
to different parameter uncertainties. Thus, the resulting multi-agent systems are weakly heterogeneous,
which fits well into the gap between the commonly-studied
homogeneous multi-agent systems and the heterogeneous multi-agent systems as
investigated in \cite{lestas2006scalable,lestas2007scalable,jonsson2010scalable}.
Typical examples belonging to this scenario are the mass-spring
systems \cite{ren2008synchronization}
with different or uncertain spring constants, the Lorenz-type chaotic systems
that cover the Lorenz, Chen, L\"{u} systems as special cases with the change
of a key parameter \cite{duan2009global}, and the discrete-time double integrators
with unknown model parameters
which have applications in synchronization of a network of clocks \cite{carli2011distributed}.

In this paper, distributed controllers are proposed
for both the continuous- and discrete-time uncertain multi-agent systems, which rely
on the relative states between neighboring agents and the absolute states of a subset of the agents.
It is shown for both the continuous- and discrete-time cases that the distributed robust control problems under
such controllers in the sense of quadratic stability
(which are referred to as distributed quadratic stabilization problems)
are equivalent to the $H_\infty$ control problems
of a set of decoupled linear systems having the same dimensions as a single agent.
A two-step algorithm is presented to construct
the distributed controller for the continuous-time case, which decouples
the feedback gain design from the communication topology and does not involve any conservatism.
The maximal allowable uncertainty bound is given as a corollary. A sufficient existence condition
in terms of linear matrix inequalities is further derived for the distributed discrete-time controller.
Note that
such a condition is conservative to some extent, because the eigenvalues of the stochastic matrix
associated with the communication graph are treated as an uncertainty.
Moreover, distributed quadratic stabilization
with $H_\infty$ disturbance attenuation
is considered for uncertain linear multi-agent systems subject to external disturbances,
which can be reduced to
the scaled $H_\infty$ control problems of a set of independent
systems whose dimensions are equal to that of a single agent. Design
procedures for distributed controllers are further
given for both the continuous- and discrete-time cases. 

The rest of this paper is organized as follows. Some basic notation
and useful results of the graph theory are reviewed in Section 2.
The distributed robust control
problems of continuous- and discrete-time multi-agent
systems are investigated in, respectively, Sections 3 and 4.
Simulation examples are given for illustration in
Section 5. Conclusions are drawn in Section 6.

\section{Graph Theory and Notation}

Let $\mathbf{R}^{n\times n}$ be the set of $n\times n$ real
matrices. The superscript $T$ means the transpose for real matrices.
$I_N$ represents the identity matrix of dimension $N$. Matrices, if
not explicitly stated, are assumed to have compatible dimensions.
Denote by $\mathbf{1}$ the column vector with all entries equal to
one. $\rm{diag}(A_1,\cdots,A_n)$ represents a block-diagonal matrix with
matrices $A_i,i=1,\cdots,n,$ on its diagonal. The matrix inequality
$A>B$ (respectively, $A\geq B$) means that $A-B$ is positive
definite (respectively, positive semi-definite). $A\otimes B$
denotes the Kronecker product of matrices $A$ and $B$.
A matrix is Hurwitz (in the continuous-time
case) if all of its eigenvalues have negative real parts,
while is Schur stable (in the discrete-time case) if all of
its eigenvalues have magnitude less than 1.

An undirected graph $\mathcal {G}$ is a pair $(\mathcal {V},\mathcal
{E})$, where $\mathcal {V}=\{v_1,\cdots,v_m\}$ is the set of nodes
and $\mathcal {E}\subseteq\mathcal {V}\times\mathcal {V}$ is the set
of unordered pairs of nodes, called edges. Two nodes $v_i$, $v_j$
are adjacent, or neighboring, if $(v_i,v_j)$ is an edge of graph
$\mathcal {G}$. A path on $\mathcal {G}$ from node $v_{i_1}$ to node
$v_{i_l}$ is a sequence of ordered edges of the form $(v_{i_k},
v_{i_{k+1}})$, $k=1,\cdots,l-1$. A graph is called connected if
there exists a path between every pair of distinct nodes, otherwise
is disconnected.

The adjacency matrix $\mathcal {A}=[a_{ij}]\in\mathbf{R}^{m\times
m}$ associated with the undirected graph $\mathcal {G}$ is defined
by $a_{ii}=0$, $a_{ij}=a_{ji}=1$ if $(i,j)\in\mathcal {E}$, and
$a_{ij}=a_{ji}=0$ otherwise. The Laplacian matrix $\mathcal
{L}\in\mathbf{R}^{m\times m}$ is defined as $\mathcal
{L}_{ii}=\sum_{j\neq i}a_{ij}$ and $\mathcal {L}_{ij}=-a_{ij}$,
$i\neq j$. Let $\mathcal {D}=[d_{ij}]\in\mathbf{R}^{m\times m}$ be a
double-stochastic matrix associated with $\mathcal {G}$ with the
additional assumption that $d_{ii}>0$, $d_{ij}=d_{ji}>0$ if
$(i,j)\in\mathcal {E}$ and $d_{ij}=d_{ji}=0$ otherwise. It is
straightforward to verify that zero is an eigenvalue of $\mathcal
{L}$ with $\mathbf{1}$ as the corresponding eigenvector and all
nonzero eigenvalues are positive. One is an eigenvalue
of $\mathcal {D}$ with $\mathbf{1}$ as the corresponding eigenvector
and all other eigenvalues of $\mathcal {D}$ are in the open unit
disk \cite{ren2005consensus}.

\section{Distributed Robust Control of Uncertain Continuous-Time Multi-Agent Systems}

\subsection{Distributed Quadratic Stabilization}

Consider a network consisting of $N$ continuous-time linear agents subject to
parameter uncertainties, described by
\begin{equation}\label{1}
\begin{aligned}
\dot{x}_i &=(A+\Delta A_i)x_i+Bu_i,\qquad i=1,\cdots,N,
\end{aligned}
\end{equation}
where $x_i\in\mathbf{R}^n$ and $u_i\in\mathbf{R}^m$ are,
respectively, the state and the control input of the $i$-th agent,
$A$ and $B$ are constant matrices with compatible
dimensions, and $\Delta A_i$ is an
unknown matrix which represents the time-varying uncertainty
associated with the $i$-th agent and is assumed to be in the form
of $\Delta A_i=DF_iE$, where $F_i\in\mathbf{R}^{j\times k}$ is the
uncertainty satisfying
\begin{equation}\label{uncer}
\begin{aligned}
F_i^TF_i\leq \delta^2I,\qquad i=1,\cdots,N,
\end{aligned}
\end{equation}
with elements of $F_i$ being Lebesgue measurable and $\delta>0$ a
given constant, and $D$ and $E$ are known constant matrices which
characterize the structure of the uncertainty.

The communication topology among the $N$ agents is represented by an
undirected graph $\mathcal {G}$.
It is assumed here
that only a subset of agents know their own states but each agent
can measure the relative states with respect to its neighbors. Without
loss of generality, assume that the first $q$ $(1\leq q< N)$ agents
have access to their state information.

The following assumption will be used in the sequel.

{\small \bf Assumption 1}. The undirected communication graph $\mathcal {G}$ is
connected and at least one agent knows its own state.

Based on the relative state information between neighboring agents
and the absolute states of a portion of agents, a
distributed controller is proposed as
\begin{equation}\label{cl}
\begin{aligned}
u_i =cK\left(\sum_{j=1}^Na_{ij}(x_i-x_j)+d_ix_i\right),\qquad i=1,\cdots,N,
\end{aligned}
\end{equation}
where $c>0\in\mathbf{R}$ is the coupling strength,
$K\in\mathbf{R}^{m\times n}$ is the feedback gain matrix,
$a_{ij}$ is the $(i,j)$-th entry of the adjacency matrix $\mathcal {A}$
associated with $\mathcal {G}$, and $d_i$ are constant scalars,
satisfying $d_i>0$, $i=1,\cdots,q,$ and $d_i=0$,
$i=q+1,\cdots,N$.

Let $x=[x_1^T,\cdots,x_N^T]^T$ and $\widehat{D}={\rm{diag}}(d_1,\cdots,d_N)$.
Then, the closed-loop network
dynamics resulting from \dref{1} and \dref{cl} can be rewritten as
\begin{equation}\label{net}
\begin{aligned}
\dot{x} &=\left(I_N\otimes A+c\widehat{\mathcal {L}}\otimes BK+(I_N\otimes
D)\Delta( I_N\otimes E)\right)x,
\end{aligned}
\end{equation}
where $\widehat{\mathcal {L}}=\mathcal {L}+\widehat{D}$,
$\mathcal {L}$ is the Laplacian matrix of $\mathcal {G}$, and
$\Delta={\rm{diag}}(F_1,\cdots,F_{N})$.

First, the notion of quadratic stability is introduced.

{\small \bf Definition 1} \cite{khargonekar1990robust,xie1992h}. The system
\dref{1} with $u_i=0$ is quadratically stable if there exists a
positive-definite matrix $P$ such that for all admissible
uncertainty $\Delta A_i$,
$$P(A+\Delta A_i)^T+(A+\Delta A_i)P<0.$$

The objective in this section is to design a distributed controller
\dref{cl} such that the closed-loop network \dref{net} is quadratically stable
for all admissible uncertainties $\Delta A_i$,
$i=1,\cdots,N$. This problem is referred to as distributed quadratic
stabilization problem.

{\small \bf Lemma 1} \cite{xie1992h}. The system
\dref{1} with $u_i=0$ is quadratically stable for all
admissible uncertainties $F_i$ satisfying \dref{uncer} if and only if
$A$ is Hurwitz and
$\|E(sI-A)^{-1}D\|_\infty<\frac{1}{\delta}$.

{\small \bf Lemma 2} \cite{hong2006tracking,li2010consensus}.
For a graph $\mathcal {G}$ satisfying Assumption 1,
the matrix $\widehat{\mathcal {L}}$ in \dref{net} is positive
definite.

The following presents a necessary and sufficient condition for the distributed quadratic
stabilization problem.

{\small \bf Theorem 1}. Under Assumption 1, the closed-loop network
\dref{net} is quadratically stable for all admissible
uncertainties $F_i$, $i=1,\cdots,N$, satisfying
\dref{uncer}, if and only if the matrices $A+c\lambda_iBK$ are Hurwitz and
$\|T_i(s)\|_\infty<\frac{1}{\delta}$, $i=1,\cdots,N$, where
$T_i(s)=E(sI-A-c\lambda_iBK)^{-1}D$, $i=1,\cdots,N$, and
$\lambda_i$, $i=1,\cdots,N$, are the eigenvalues of
$\widehat{\mathcal {L}}$.

{\small \bf Proof}.
(Necessity) Consider a special case where the certainties in \dref{1} are the same,
i.e., $F_1=\cdots=F_{N}=F$. Thus, the system \dref{net}
can be rewritten as
\begin{equation}\label{net2}
\begin{aligned}
\dot{x} &=\left(I_N\otimes A+c\widehat{\mathcal {L}}\otimes BK+I_N\otimes
DFE\right)x.
\end{aligned}
\end{equation}
Because Assumption 1 holds, it follows
from Lemma 2 that $\widehat{\mathcal {L}}$ is positive definite. Let
$U\in\mathbf{R}^{N\times N}$ be such a unitary matrix that
$U^T\widehat{\mathcal {L}}U=\Lambda\triangleq
{\rm{diag}}(\lambda_1,\cdots,\lambda_{N})$.
Let $\xi\triangleq[\xi_1^T,\cdots,\xi_{N}^T]^T=(U\otimes I_n)\xi$.
Then, it follows from \dref{net2} that
\begin{equation}\label{net3}
\begin{aligned}
\dot{\xi} &=(I_N\otimes A+c\Lambda\otimes BK+I_N\otimes
DFE)\xi.
\end{aligned}
\end{equation}
Note that the state matrix of \dref{net3} is block diagonal.
Therefore, the network \dref{net2} is quadratically stable
if and only if the following $N$ systems:
\begin{equation}\label{th1}
\begin{aligned}
\dot{\xi}_i =(A+c\lambda_iBK+DFE)\xi_i,\quad i=1,\cdots,N,
\end{aligned}
\end{equation}
are simultaneously quadratically stable,
which by Lemma 1 implies that $A+c\lambda_iBK$ are Hurwitz and
$\|T_i(s)\|_\infty<\frac{1}{\delta}$, $i=1,\cdots,N$.


(Sufficiency) From \dref{uncer}, it follows that the
uncertainty $\Delta$ in \dref{net} satisfies
$\Delta^T\Delta\leq\delta^2I$. 
In light of Lemma 1, the system \dref{net} is quadratically stable
for all admissible uncertainties $F_i$, $i=1,\cdots,N$, satisfying
\dref{uncer}, if
$I_N\otimes A+c\widehat{\mathcal {L}}\otimes BK$ is Hurwitz and
$\|T(s)\|_\infty<\frac{1}{\delta}$, where
\begin{equation}\label{tran}
\begin{aligned}
T(s) \triangleq (I_N\otimes E)\left(sI-I_N\otimes
A-c\widehat{\mathcal {L}}\otimes BK\right)^{-1}(I_N\otimes D).
\end{aligned}
\end{equation}
Note that
\begin{equation}\label{th12}
\begin{aligned}
(U\otimes I_n)T(s)(U^T\otimes I_n)&= (I_N\otimes
E)\left(sI-I_N\otimes
A-cU^T\widehat{\mathcal {L}}U\otimes BK\right)^{-1}(I_N\otimes D)\\
&= (I_N\otimes E)\left(sI-I_N\otimes A-c\Lambda\otimes BK\right)^{-1}(I_N\otimes D)\\
&={\rm{diag}}\left(T_1(s),\cdots,T_{N}(s)\right).
\end{aligned}
\end{equation}
By the definition of the $H_\infty$ norm \cite{zhou1998essentials},
it follows readily from \dref{th12} that
$$\|T(s)\|_\infty=\underset{i=1,\cdots,N}\max\|T_i(s)\|_\infty.$$
Therefore, if the matrices $A+c\lambda_iBK$ are Hurwitz and
$\|T_i(s)\|_\infty<\frac{1}{\delta}$, $i=1,\cdots,N$,
then the system \dref{net} is quadratically stable
for all admissible uncertainties $F_i$, $i=1,\cdots,N$, satisfying
\dref{uncer}.
\hfill $\blacksquare$

{\small \bf Remark 1}. Although the uncertainty $\Delta$ in \dref{net}
is structural, it is shown in Theorem 1 that the distributed
quadratic stabilization problem of \dref{net}
is equivalent to the $H_\infty$ control problems
of a set of decoupled linear systems having the same dimensions as a single agents,
which is essentially due to the fact that the nominal dynamics
of the agents are identical.

{\small \bf Remark 2}.
Different from most of the existing references focusing on homogeneous multi-agent systems,
the current paper considers the uncertain multi-agent systems where
the agents have the same nominal dynamics but subject to different parameter uncertainties.
The resulting agent network \dref{net} is thus weakly
heterogeneous, which fits well into the gap between
the commonly-studied homogeneous multi-agent systems and the heterogeneous multi-agent systems
as investigated in
\cite{lestas2006scalable,lestas2007scalable,jonsson2010scalable,wang2010distributed}.
Typical examples belonging to this scenario are the mass-spring
systems \cite{ren2008synchronization} with different or uncertain spring constants,
the Lorenz-type chaotic systems \cite{duan2009global}, and the
discrete-time double integrators with unknown model parameters \cite{carli2011distributed}.

Next, an algorithm is presented to determine the distributed
controller \dref{cl}.

{\small \bf Algorithm 1}. Under Assumption 1, a distributed controller
\dref{cl} can be constructed as follows:
\begin{itemize}
\item[1)] Solve the following linear matrix inequality (LMI):
\begin{equation}\label{tp21}
    \begin{bmatrix} AP+PA^T-\tau BB^T & \delta D & PE^T
    \\\delta D^T & -I & 0 \\
    EP & 0 & -I \end{bmatrix}<0
\end{equation}
to get a matrix $P>0$ and a scalar $\tau>0$. Then, choose $K=-\frac{1}{2}B^TP^{-1}$.

\item[2)]
Select the coupling strength $c\geq c_{th}$, where
\begin{equation}\label{cupc}
c_{th}=\frac{\tau}{\underset{i=1,\cdots,N}{\min}(\lambda_i)}.
\end{equation}
\end{itemize}

{\small \bf Theorem 2}. Under Assumption 1, the closed-loop network
\dref{net} with the distributed controller constructed by Algorithm 1
is quadratically stable for all admissible
uncertainties $F_i$, $i=1,\cdots,N$, satisfying \dref{uncer}.

{\small \bf Proof}. By the Schur Complement Lemma
\cite{boyd1994linear}, the LMI \dref{tp21} is feasible for some
matrix $P>0$ and scalar $\tau>0$ if and only if the following
Riccati inequality holds:
\begin{equation}\label{p21}
AP+PA^T-\tau BB^T+\delta^2DD^T+PE^TEP<0.
\end{equation}
From step 2) in Algorithm 1, it follows that $c\lambda_i\geq \tau$, $i=1,\cdots,N$.
Noting $K=-\frac{1}{2}B^TP^{-1}$,
it can be obtained that
\begin{equation}\label{st2s}
\begin{aligned}
(A+c\lambda_iBK)P&+P(A+c\lambda_iBK)^T+\delta^2DD^T+PE^TEP\\
&=AP+PA^T-c\lambda_i BB^T+\delta^2DD^T+PE^TEP\\
&<0.
\end{aligned}
\end{equation}
By the Bounded Real Lemma \cite{zhou1998essentials}, it follows from
\dref{st2s} that $A+c\lambda_iBK$ are Hurwitz and
$\|T_i(s)\|_\infty<\frac{1}{\delta}$, $i=1,\cdots,N$. Therefore,
Theorem 1 implies that the network \dref{net} in this case is
quadratically stable for all admissible uncertainties
$F_i$, $i=1,\cdots,N$, satisfying \dref{uncer}. \hfill $\blacksquare$

{\small \bf Remark 3}. As shown in the proof of Proposition 1 in
\cite{li2010consensus}, by using Finsler's Lemma
\cite{iwasaki1994all}, it is not difficult to see that there exist
$P>0$ and $\tau>0$ such that \dref{tp21} holds if and only if
there exists a $K$ such that
$(A+BK)P+P(A+BK)^T+\delta^2DD^T+PE^TEP<0$. That is, $A+BK$ is Hurwitz
and $\|E(sI-A-BK)^{-1}D\|_\infty<\frac{1}{\delta}$.
Therefore, it follows from Theorems 1 and 2 that
the feasibility
of the LMI \dref{tp21} with respect to $P>0$ and $\tau>0$
is not only sufficient but also necessary for
the existence of a controller \dref{cl} satisfying Theorem 1.
The largest allowable uncertainty bound
$\delta_{\max}$ can be obtained by the following optimization
problem:
\begin{equation}\label{cor1}
\begin{aligned}
&\rm{maximize}\quad\delta>0 \\
&\text{subject to\quad LMI} ~\dref{tp21},~\text{with}~P>0,~\tau>0.
\end{aligned}
\end{equation}

\subsection{Distributed Quadratic $H_\infty$ Control}

This subsection extends to consider a network of uncertain linear agents
subject to external disturbance, given by
\begin{equation}\label{d1}
\begin{aligned}
\dot{x}_i &=(A+\Delta A_i)x_i+Bu_i+B_2\omega_i,\\
z_i &= Cx_i,\qquad i=1,\cdots,N,
\end{aligned}
\end{equation}
where $\omega_i\in\mathbf{R}^{p}$ and $z_i\in\mathbf{R}^l$ are, respectively,
the exogenous disturbance and the performance variable of the $i$-th agent,
and the rest of the variables are defined as in \dref{1}.

The distributed controller is still given as in \dref{cl}. Let
$\omega=[\omega_1^T,\cdots,\omega_N^T]^T$ and
$z=[z_1^T,\cdots,z_N^T]^T$. Then, the closed-loop system resulting
from \dref{d1} and \dref{cl} can be written as
\begin{equation}\label{netd}
\begin{aligned}
\dot{x}  &= \left(I_N\otimes A+c\widehat{\mathcal {L}}\otimes BK+(I_N\otimes
D)\Delta( I_N\otimes E)\right)x
+(I_N\otimes B_2)\omega, \\
z & = (I_N\otimes C)x.
\end{aligned}
\end{equation}

This section is to design a distributed controller \dref{cl} such
hat the closed-loop system \dref{netd} is quadratically stable and
meanwhile
achieves a prescribed level of disturbance attenuation in the
$H_\infty$-norm sense for all admissible uncertainties $F_i$,
$i=1,\cdots,N$, satisfying \dref{uncer}.

{\small \bf Definition 2} \cite{xie1992h}. The system \dref{d1} with
$u_i=0$ is quadratically stable with a disturbance attenuation level
$\gamma>0$ if there exists a positive-definite matrix $P$ such that
for all admissible uncertainty $\Delta A_i$,
$$P(A+\Delta A_i)^T+(A+\Delta A_i)P+\frac{1}{\gamma^2}B_2B_2^T+PC^TCP<0.$$

{\small \bf Lemma 3} \cite{xie1992h}. There exists a
positive-definite matrix $P$ such that
$$P(A+H_1F_1E_1)^T+(A+H_1F_1E_1)P<0$$
for all admissible uncertainty $F_1(t)$ satisfying $F_1^TF_1\leq
\varrho^2I$ if and only if there exists a scalar $\epsilon>0$ such
that
$$PA^T+AP+\frac{1}{\epsilon}PE_1^TE_1P+\epsilon\varrho^2H_1H_1^T<0.$$

{\small \bf Theorem 3}. Suppose that Assumption 1 holds. Then, the closed-loop network
\dref{netd} is quadratically stable with disturbance attenuation
$\gamma>0$ for all admissible uncertainties $F_i$,
$i=1,\cdots,N$, satisfying \dref{uncer}, if the following systems:
\begin{equation}\label{th31}
\begin{aligned}
\dot{x}_i &=(A+c\lambda_iBK)x_i+\begin{bmatrix}
\epsilon^{\frac{1}{2}}\delta D &
\gamma^{-1}B_2\end{bmatrix}\bar{\omega}_i,\\
\bar{z}_i&=\begin{bmatrix} \epsilon^{-\frac{1}{2}} E \\
C\end{bmatrix}x_i, \quad i=1,\cdots,N,
\end{aligned}
\end{equation}
are simultaneously asymptotically stable with unitary disturbance attenuation,
where $x_i\in\mathbf{R}^n$, $\bar{\omega}_i\in\mathbf{R}^{p+j}$,
$\bar{z}_i\in\mathbf{R}^{l+k}$, and $\epsilon>0$ is a scalar to be
chosen.

{\small \bf Proof}. As stated in the proof of Theorem 1, the
uncertainty $\Delta$ in \dref{netd} is structural and satisfies
$\Delta^T\Delta\leq\delta^2I$. In view of Definition 2, the network \dref{netd}
is quadratically stable with disturbance attenuation $\gamma>0$ for
all admissible uncertainties $F_i$, $i=1,\cdots,N$,
satisfying \dref{uncer}, if there exists a
positive-definite matrix $\mathcal {P}\in\mathbf{R}^{Nn\times Nn}$
such that
\begin{equation}\label{th32}
\begin{aligned}
\mathcal {P}\mathcal {S}^T+\mathcal {S}\mathcal
{P}+\frac{1}{\gamma^2}\mathcal {B}_2\mathcal {B}_2^T+ \mathcal
{P}\mathcal {C}^T\mathcal {C}\mathcal {P}<0,
\end{aligned}
\end{equation}
where $$\begin{aligned}
\mathcal {S} &=I_N\otimes A+c\widehat{\mathcal {L}}\otimes BK+(I_N\otimes
D)\Delta( I_N\otimes E),\\
\mathcal {B}_2 &=I_N\otimes B_2,\quad \mathcal {C} =I_N\otimes C.
\end{aligned}$$
Let $U\in\mathbf{R}^{N\times N}$ be such a unitary matrix that
$U^T\widehat{\mathcal
{L}}U=\Lambda={\rm{diag}}(\lambda_1,\cdots,\lambda_{N})$.
Multiplying the left and right sides of \dref{th32} by $U^T\otimes
I_n$ and $U\otimes I_n$, respectively, gives
\begin{equation}\label{th33}
\begin{aligned}
\widetilde{\mathcal {P}}\widetilde{\mathcal
{S}}^T+\widetilde{\mathcal {S}}\widetilde{\mathcal {P}}
+\frac{1}{\gamma^2}\mathcal {B}_2\mathcal {B}_2^T+ \widetilde{\mathcal
{P}}\mathcal {C}^T\mathcal {C}\widetilde{\mathcal {P}}<0,
\end{aligned}
\end{equation}
where $$\begin{aligned}
\widetilde{\mathcal {S}} &=I_N\otimes A+c\Lambda\otimes BK+(I_N\otimes
D)\widetilde{\Delta}( I_N\otimes E),\\
\widetilde{\Delta} &=(U^T\otimes I_n)\Delta(U\otimes I_n),
\quad\widetilde{\mathcal {P}} =(U^T\otimes I_n)\mathcal {P}(U\otimes
I_n).
\end{aligned}$$
Clearly, it follows from \dref{uncer} that $\widetilde{\Delta}$
satisfies
$$\widetilde{\Delta}^T\widetilde{\Delta}=(U^T\otimes I_n)\Delta^T\Delta(U\otimes I_n)\leq\delta^2I.$$ By
Lemma 3, it follows that \dref{th33} holds if there exists a scalar
$\epsilon$ such that
\begin{equation}\label{th34}
\begin{aligned}
\widetilde{\mathcal {P}}(I_N\otimes A+c\Lambda\otimes
BK)^T+(I_N\otimes A+c\Lambda\otimes BK)\widetilde{\mathcal {P}}
+\frac{1}{\epsilon}\widetilde{\mathcal {P}}(I_N\otimes
E^TE)\widetilde{\mathcal {P}}\\+\epsilon\delta^2I_N\otimes DD^T
+\frac{1}{\gamma^2}\mathcal {B}_2\mathcal {B}_2^T+ \widetilde{\mathcal
{P}}\mathcal {C}^T\mathcal {C}\widetilde{\mathcal {P}}<0.
\end{aligned}
\end{equation}
Note that all the matrices in \dref{th34} except
$\widetilde{\mathcal {P}}$ are block diagonal. Therefore, if there
exist positive-definite matrices $P_i\in\mathbf{R}^{n\times n}$ such
that
\begin{equation}\label{th35}
\begin{aligned}
P_i(A+c\lambda_iBK)^T+(A+c\lambda_iBK)P_i+\frac{1}{\epsilon}P_iE^TEP_i+\epsilon\delta^2DD^T\\
+\frac{1}{\gamma^2}B_2B_2^T+P_iC^TCP_i<0,\quad i=1,\cdots,N,
\end{aligned}
\end{equation}
then the matrix $\mathcal {P}=(U\otimes
I_n){\rm{diag}}(P_1,\cdots,P_N)(U^T\otimes I_n)$ satisfies
\dref{th32}. By the Bounded Real Lemma \cite{zhou1998essentials},
it is easy to see that \dref{th35} is
equivalent to that the $N$ systems in \dref{th31} are simultaneously
asymptotically stable with unitary disturbance attenuation.
This completes the proof.
\hfill $\blacksquare$

{\small \bf Remark 4}. Theorem 3 casts the
robust $H_\infty$ control problem of high-dimensional network \dref{netd} to the scaled $H_\infty$
control problems of a set of independent low-dimensional linear
systems in \dref{th31}, thereby reducing the computational complexity
significantly. When the external disturbances $\omega_i$ do not exist, Theorem 3 is reduced to Theorem 1.

{\small \bf Algorithm 2}. Under Assumption 1, a distributed controller
\dref{cl} can be constructed as follows:
\begin{itemize}
\item[1)] Solve the following LMI:
\begin{equation}\label{tp41}
    \begin{bmatrix} AQ+QA^T-\tilde{\tau} BB^T+\frac{1}{\gamma^2}B_2B_2^T+\epsilon\delta^2 DD^T & QC^T
    &  QE^T    \\ CQ & -I & 0 \\
    EQ & 0 & -\epsilon I \end{bmatrix}<0
\end{equation}
to get a matrix $Q>0$ and scalars $\tilde{\tau}>0$, $\epsilon>0$. Then,
choose $K=-\frac{1}{2}B^TQ^{-1}$.

\item[2)]
Select the coupling strength $c\geq c_{th}$, where $c_{th}$ is
defined in \dref{cupc}.
\end{itemize}

{\small \bf Remark 5}. It is worth
noting that in Algorithms 1 and 2, the feedback gain design of \dref{cl}
is decoupled from the communication topology
and only the smallest eigenvalue of $\widehat{\mathcal {L}}$ is used
to select the coupling strength $c$.
One consequence of this decoupling property is that
the controller designed for one given
connected communication graph can be used directly to any other connected graphs, with the
only task of appropriately selecting the coupling strength $c$.
By selecting the coupling strength $c$ to be relatively large, the distributed
controller \dref{cl} constructed by these algorithms maintains certain degree of robustness with respect to variations
of the communication graph $\mathcal {G}$, such as adding or removing edges or agents in $\mathcal {G}$,
in which case although
the eigenvalues $\lambda_i$, $i=1,\cdots,N$, are changed,
$c\lambda_i$, $i=1,\cdots,N$, can still be larger than $\tau$ (or $\tilde{\tau}$).

{\small \bf Theorem 4}. Suppose that Assumption 1 holds. Then, the closed-loop network
\dref{netd} with the distributed controller \dref{cl} constructed by
Algorithm 2 is quadratically stable with disturbance attenuation
$\gamma>0$ for all admissible uncertainties $\Delta A_i$,
$i=1,\cdots,N$, satisfying \dref{uncer}

{\small \bf Proof}.
The proof is similar to that of Theorem 2 and thus is omitted.
\hfill $\blacksquare$

{\small \bf Corollary 1}. As $\gamma\rightarrow\infty$, the LMI
condition \dref{tp41} in Algorithm 2 is reduced to \dref{tp21} in
Algorithm 1.

{\small \bf Proof}.
By the Schur Complement lemma \cite{boyd1994linear}, the LMI \dref{tp41} is equivalent to
$$
\begin{aligned}
QA^T+AQ-\tilde{\tau} BB^T+\frac{1}{\gamma^2}B_2B_2^T+\epsilon\delta^2
DD^T+\frac{1}{\epsilon}QE^TEQ+QC^TCQ<0.
\end{aligned}
$$
Letting $Q=\gamma P$, $\epsilon=\gamma^{-1}$, and $\tilde{\tau}=\gamma^{-1}\tau$ gives
\begin{equation}\label{th452}
\begin{aligned}
PA^T+AP-\tau BB^T+\frac{1}{\gamma}B_2B_2^T+\delta^2
DD^T+PE^TEP+\frac{1}{\gamma}PC^TCP<0.
\end{aligned}
\end{equation}
When $\gamma\rightarrow\infty$, \dref{th452} implies that
$$
\begin{aligned}
PA^T+AP-\tau BB^T+\delta^2 DD^T+PE^TEP<0,
\end{aligned}
$$
which is equivalent to \dref{tp21}.
\hfill $\blacksquare$

\section{Distributed Robust Control of Uncertain Discrete-Time Multi-Agent Systems}

This section considers a network consisting of $N$ discrete-time
agents with parameter uncertainties, described by
\begin{equation}\label{di1}
\begin{aligned}
x_i(k+1) &=(A+\Delta A_i)x_i(k)+Bu_i(k),\qquad i=1,\cdots,N,
\end{aligned}
\end{equation}
where $x_i(k)\in\mathbf{R}^n$ and $u_i(k)\in\mathbf{R}^m$ are,
respectively, the state and the control input at the $k$ time instant of the $i$-th agent,
and $\Delta A_i$ denotes the time-varying uncertainty
associated with the $i$-th agent, which is assumed to be in the form
of $\Delta A_i=DF_iE$, with $F_i\in\mathbf{R}^{j\times k}$ satisfying
\dref{uncer}.

Similar to \dref{cl}, a
distributed controller based on the relative states between neighboring agents
and the absolute states of a subset of agents is proposed as
\begin{equation}\label{cldi}
\begin{aligned}
u_i(k) =K\left(\sum_{j=1}^Ne_{ij}(x_i(k)-x_j(k))+\hat{d}_ix_i(k)\right),\qquad i=1,\cdots,N,
\end{aligned}
\end{equation}
where $K\in\mathbf{R}^{m\times n}$ is the feedback gain matrix,
$e_{ij}$ is the $(i,j)$-th entry of
the double-stochastic matrix $\mathcal {D}$ associated with $\mathcal {G}$,
and $\hat{d}_i$ are constant scalars,
satisfying $e_{ii}>\hat{d}_i>0$, $i=1,\cdots,q,$ and $\hat{d}_i=0$,
$i=q+1,\cdots,N$.

Let $x=[x_1^T,\cdots,x_N^T]^T$ and $\widehat{D}={\rm{diag}}(\hat{d}_1,\cdots,\hat{d}_N)$.
Then, the closed-loop network
dynamics resulting from \dref{di1} and \dref{cldi} can be rewritten as
\begin{equation}\label{netdi}
\begin{aligned}
x(k+1) &=\left(I_N\otimes A+(I_N-\widetilde{\mathcal {D}})\otimes BK+(I_N\otimes
D)\Delta( I_N\otimes E)\right)x(k),
\end{aligned}
\end{equation}
where $\widetilde{\mathcal {D}}=\mathcal {D}-\widehat{D}$ and
$\Delta={\rm{diag}}(F_1,\cdots,F_{N})$.

The objective in this section is to solve the discrete-time distributed
quadratic stabilization problem for \dref{netdi},
i.e., to design a distributed controller
\dref{cldi} such that the closed-loop network \dref{netdi} is quadratically stable
for all admissible uncertainties $\Delta A_i$,
$i=1,\cdots,N$.

The notion of quadratic stability for \dref{di1} is introduced below.

{\small \bf Definition 3} \cite{de1993h}.
The system
\dref{di1} with $u_i=0$ is quadratically stable if there exists a
positive-definite matrix $P$ such that for all admissible
uncertainty $\Delta A_i$,
$$(A+\Delta A_i)^TP(A+\Delta A_i)-P<0.$$

{\small \bf Lemma 4} \cite{de1993h}.
The system
\dref{di1} with $u_i=0$ is quadratically stable for all
admissible uncertainties $F_i$ satisfying \dref{uncer} if and only if
$A$ is Schur stable and
$\|E(zI-A)^{-1}D\|_\infty<\frac{1}{\delta}$.

{\small \bf Lemma 5}. Suppose that
the graph $\mathcal {G}$ satisfies Assumption 1.
Then, the matrix $\widetilde{\mathcal {D}}$ in \dref{netdi}
have all of its eigenvalues located inside the unit circle.

{\small \bf Proof}. Consider the following new row-stochastic
matrix:
$$\overline{\mathcal {D}}=\begin{bmatrix}\widetilde{\mathcal {D}} & \widehat{D}{\bf 1}_{N}\\
0_{1\times N} & 1\end{bmatrix}.$$
According to the definition of the
directed spanning tree \cite{ren2005consensus}, the graph associated with
$\overline{\mathcal {D}}$ has a directed spanning tree if $\mathcal
{G}$ satisfies Assumption 1. Therefore, by Lemma 3.4 in
\cite{ren2005consensus}, if Assumption 1 holds, then 1 is a simple
eigenvalue of $\overline{\mathcal {D}}$ and all the other
eigenvalues of $\overline{\mathcal {D}}$ are in the open unit disk,
which further implies that all the eigenvalues of $\widetilde{\mathcal {D}}$
lie in the open unit disk.\hfill $\blacksquare$

The following gives a sufficient condition for solving the distributed quadratic
control problem of \dref{netdi}.

{\small \bf Theorem 5}. Under Assumption 1, the closed-loop network
\dref{netdi} is quadratically stable for all admissible
uncertainties $F_i$, $i=1,\cdots,N$, satisfying
\dref{uncer}, if and only if the matrices $A+(1-\tilde{\lambda}_i)BK$ are Schur stable and
$\|\tilde{T}_i(z)\|_\infty<\frac{1}{\delta}$, $i=1,\cdots,N$, where
$\tilde{T}_i(z)=E(zI-A-(1-\tilde{\lambda}_i)BK)^{-1}D$, $i=1,\cdots,N$, and
$\tilde{\lambda}_i$, $i=1,\cdots,N$, are the eigenvalues of
$\widetilde{\mathcal {D}}$.

Furthermore, if there exist matrices $Q>0$, $W$, and a scalar $\tau$ such that
\begin{equation}\label{th5}
\begin{bmatrix}-Q & (AQ+BW)^T & QE^T & W^T\\AQ+BW & -Q+\delta^2 DD^T+\tau \kappa^2 BB^T & 0 & 0\\
EQ & 0 & -I &0\\
W & 0 & 0& -\tau I\end{bmatrix}<0,
\end{equation}
where $\kappa=\underset{i=1,\cdots,N}\max |\tilde{\lambda}_i|$, then there exists
a distributed controller \dref{cldi} such that \dref{netdi} is quadratically stable.
Specifically, the feedback gain matrix of
\dref{cldi} is given by $K=WQ^{-1}$.

{\small \bf Proof}. The equivalence between the quadratic stability
of \dref{netdi} and $\|\tilde{T}_i(z)\|_\infty<\frac{1}{\delta}$ can be checked
by following similar steps in the proof of Theorem 1. 
Next, it will be shown that the feasibility of the LMI \dref{th5} guarantees the existence of
a controller \dref{cldi} satisfying the quadratic stability of \dref{netdi}. In
virtue of the discrete-time Bounded Real Lemma \cite{zhou1998essentials},
the matrices $A+(1-\tilde{\lambda}_i)BK$ are Schur stable and
$\|\tilde{T}_i(z)\|_\infty\leq\frac{1}{\delta}$, $i=1,\cdots,N$, if and only if
there exist matrices $P_i>0$ such that
\begin{equation}\label{th52}
\hat{A}^TP_i \hat{A}-P_i+E^TE+\delta^2\hat{A}^TP_i D(I-\delta^2 DP_iD^T)^{-1}DP_i\hat{A}<0,\quad i=1,\cdots,N,
\end{equation}
where $\hat{A}=A+(1-\tilde{\lambda}_i)BK$. Noting that $|\tilde{\lambda}_i|\leq \kappa$, $i=1,\cdots,N$,
\dref{th52} clearly hold for $i=1,\cdots,N$, if there exists a $P>0$ such that
\begin{equation}\label{th521}
\tilde{A}^TP \tilde{A}-P+E^TE+\delta^2\tilde{A}^TP D(I-\delta^2 DPD^T)^{-1}DP\tilde{A}<0,
\end{equation}
where $\tilde{A}=A+(1-\tilde{\Delta})BK$, for all $|\tilde{\Delta}|\leq \kappa$. It is not difficult
to see that \dref{th521} can be rewritten as
$$
\tilde{A}^T(P^{-1}- \delta^2DD^T)^{-1}\tilde{A} -P+E^TE<0.
$$
By letting $Q=P^{-1}$ and applying the Schur Complement Lemma \cite{boyd1994linear}, the above inequality
is equivalent to
$$
\begin{bmatrix}-Q^{-1}& \tilde{A}^T & E^T\\
\tilde{A} & Q-\delta^2 DD^T &0 \\
E & 0 & -I\end{bmatrix}<0,
$$
which can be rewritten as
\begin{equation}\label{th54}
\begin{bmatrix}-Q^{-1}& (A+BK)^T & E^T\\
A+BK & Q-\delta^2 DD^T &0\\
E & 0 & -I\end{bmatrix}+\begin{bmatrix} 0 \\ B\\0\end{bmatrix}
\tilde{\Delta}\begin{bmatrix} K & 0 &0\end{bmatrix}+\begin{bmatrix} K^T \\ 0\\0\end{bmatrix}
\tilde{\Delta}\begin{bmatrix} 0 & B^T&0\end{bmatrix}<0.
\end{equation}
Using Lemma 3, \dref{th54} holds if and only if
\begin{equation}\label{th552}
\begin{bmatrix}-Q^{-1}& (A+BK)^T & E^T& K^T \\
A+BK & Q-\delta^2 DD^T+ \tau \kappa^2 BB^T &0 &0\\
E & 0 & -I &0\\
K & 0 & 0& -\tau I\end{bmatrix}<0,
\end{equation}
for some scalar $\tau$. Multiplying the both sides of \dref{th552}
by ${\rm{diag}}(Q,I,I,I)$ and noting $W=KQ$ leads directly to \dref{th5}.
\hfill $\blacksquare$

{\small \bf Remark 6}. By similar steps in proving Corollary 1,
it is not difficult to show that the LMI condition \dref{th5} is equivalent to
$\|K(zI-A-BK)^{-1}B\|_\infty<\frac{1}{\kappa}$,
when $\delta\rightarrow 0$, i.e., the uncertainty bound is sufficiently
small. As pointed out in \cite{de1993h,fu2005sector},
a sufficient condition for the existence of $K$ satisfying $\|K(zI-A-BK)^{-1}B\|_\infty<\frac{1}{\kappa}$
is that $(A,B)$ is stabilizable and $\kappa\leq\frac{1}{\Pi_i|\lambda_i^u(A)|}$, where
$\lambda_i^u(A)$ are the unstable eigenvalues of $A$. Therefore, under such a condition,
the LMI \dref{th5} is feasible for the case where
the uncertainty bound in \dref{uncer} is sufficiently small. The largest allowable uncertainty bound
$\delta_{\max}$ can be obtained by maximizing $\delta$ in \dref{th5}.

{\small \bf Remark 7}. Theorem 5 involves certain conservatism, which is introduced by treating
the eigenvalues $\lambda_i$, $i=1,2,\cdots,N$, as uncertainties with the bound $\kappa$.
By selecting $\kappa$ to be relatively large,
the distributed controller \dref{cldi} constructed by Theorem 5
maintains certain degree of robustness with respect to variations of the communication graph $\mathcal {G}$,
such as adding or removing edges in $\mathcal {G}$.

Next, consider a network of uncertain discrete-time agents
subject to external disturbance, given by
\begin{equation}\label{did1}
\begin{aligned}
x_i(k+1) &=(A+\Delta A_i)x_i(k)+Bu_i(k)+B_2\omega_i(k),\\
z_i(k+1) &= Cx_i(k),\qquad i=1,\cdots,N,
\end{aligned}
\end{equation}
where $\omega_i\in\mathbf{R}^{p}$, $z_i\in\mathbf{R}^l$,
and the rest of the variables are defined as in \dref{di1}.

Let
$\omega=[\omega_1^T,\cdots,\omega_N^T]^T$ and
$z=[z_1^T,\cdots,z_N^T]^T$. Then, it follows from \dref{did1} and \dref{cldi}
that the closed-loop system can be written as
\begin{equation}\label{netdid}
\begin{aligned}
x(k+1)  &= \left(I_N\otimes A+(I_N-\widetilde{\mathcal {D}})\otimes BK+(I_N\otimes
D)\Delta( I_N\otimes E)\right)x(k)
+(I_N\otimes B_2)\omega(k), \\
z(k+1) & = (I_N\otimes C)x(k).
\end{aligned}
\end{equation}

{\small \bf Theorem 6}. Suppose that Assumption 1 holds. Then, the closed-loop network
\dref{netd} is quadratically stable with disturbance attenuation
$\gamma>0$ for all admissible uncertainties $F_i$,
$i=1,\cdots,N$, satisfying \dref{uncer}, if the following systems:
\begin{equation}\label{th61}
\begin{aligned}
x_i(k+1) &=(A+(1-\tilde{\lambda}_i)BK)x_i(k)+\begin{bmatrix}
\epsilon^{\frac{1}{2}}\delta D &
\gamma^{-1}B_2\end{bmatrix}\tilde{\omega}_i(k),\\
\tilde{z}_i(k+1)&=\begin{bmatrix} \epsilon^{-\frac{1}{2}} E \\
C\end{bmatrix}x_i(k), \quad i=1,\cdots,N,
\end{aligned}
\end{equation}
are simultaneously asymptotically stable with unitary disturbance attenuation,
where $x_i\in\mathbf{R}^n$, $\tilde{\omega}_i\in\mathbf{R}^{p+j}$,
$\tilde{z}_i\in\mathbf{R}^{l+k}$, and $\epsilon>0$ is a scalar to be
chosen.

Further, if there exist matrices $Q>0$, $W$, and a scalar $\tau$ such that
\begin{equation}\label{th62}
\begin{bmatrix}-Q & (AQ+BW)^T & QE^T & QC^T & W^T\\
AQ+BW & -Q+\delta^2 DD^T+\tau \kappa^2 BB^T & 0 & 0 &0\\
EQ & 0 & -\epsilon I & 0& 0\\
CQ & 0 & 0 & -I & 0\\
W & 0 & 0& 0 &-\tau I\end{bmatrix}<0,
\end{equation}
where $\kappa$ is defined as in \dref{th5}, then
the feedback gain matrix of
the controller \dref{cldi} is given by $K=WQ^{-1}$.

{\small \bf Proof}.
The proof of the first part is similar to that of Theorem 3 and the proof of the second part
is similar to that of Theorem 5. Both are omitted here for brevity.
\hfill $\blacksquare$

\section{Simulation Examples}

In this section, simulation examples are provided to validate
the effectiveness of the theoretical results.

{\small \bf Example 1}. Consider a network of mass-spring systems
with a common mass $m$ but different spring constants, described by
\begin{equation}\label{exa1}
m\ddot{y}_i+k_iy_i=u_i,\quad i=1,\cdots,N,
\end{equation}
where $y_i$ are the displacements from certain reference positions and
$k_i$, $i=1,\cdots,N,$ are the spring constants, which are assumed to be in the form of
\begin{equation}\label{exa12}
k_i=k_0+\Delta k_i,\quad i=1,\cdots,N,
\end{equation}
where $k_0$ is the identical nominal spring constant and $\Delta k_i$ are the
uncertainties satisfying $|\Delta k_i|\leq \delta$.
Denote by $x_i=\begin{bmatrix} y_i& \dot{y}_i\end{bmatrix}^T$
the state vector of the $i$-th agent. Then, \dref{exa1} and \dref{exa12} can be rewritten
as
\begin{equation}\label{exa13}
\dot{x}_i=(A+D\Delta k_iE)x_i+Bu_i,\quad i=1,\cdots,N,
\end{equation}
with $$A=\begin{bmatrix} 0 & 1\\ -\frac{k_0}{m} &0 \end{bmatrix},B=\begin{bmatrix} 0 \\ 1\end{bmatrix},
D=\begin{bmatrix} 0 \\ -\frac{1}{m} \end{bmatrix}, E=\begin{bmatrix} 1 & 0\end{bmatrix},
$$

\begin{figure}[htbp]
\centering
\includegraphics[width=0.3\linewidth]{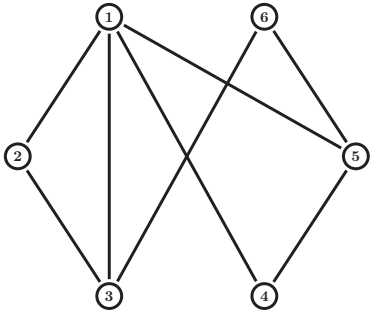}
\caption{The communication topology. }
\end{figure}

The objective is to design a distributed controller \dref{cl}
such that the closed-loop network is quadratically stable for all $\Delta k_i$.
Let $k_0=7N/m$ and $m=2.5kg$.
Solving the LMI \dref{tp21} with $\delta=10$ by using the Sedumi toolbox \cite{sturm1999using}
gives a feasible solution: $$P=\begin{bmatrix} 1.6448 & -2.3499\\ -2.3499 &
9.7007 \end{bmatrix},\quad \tau=64.0444.$$
Thus, the feedback gain matrix
is chosen as $K=-\begin{bmatrix} 0.1126 &
0.0788 \end{bmatrix}$. Assume that the communication topology is given in Fig. 1,
with only the first node knowing its own state. In \dref{cl},
let $d_1=2$, and $d_i=0$, $i=2,\cdots,6$. Then, the minimal eigenvalue
of the matrix $\widehat{\mathcal {L}}$ in \dref{net} is 0.237.
Therefore, by Algorithm 1, the controller
\dref{cl} with $K$ chosen above solves the distributed quadratic stabilization problem
for all $\Delta k_i$ satisfying $|\Delta k_i|\leq 10$, if
the coupling strength $c\geq 270.2295$. The simulation result is depicted in Fig. 2,
with $c=275$ and $\Delta k_i$ randomly chosen within the interval $(7,22]$.

\begin{figure}[htbp]
\centering
\includegraphics[width=0.5\linewidth,height=0.3\linewidth]{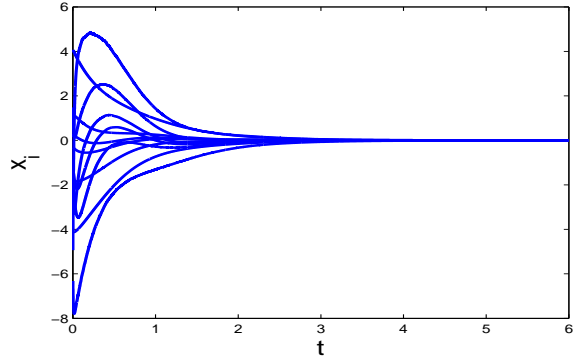}
\caption{The state trajectories of the agents \dref{exa13} under controller \dref{cl}. }
\end{figure}

By solving the optimization problem \dref{cor1},
it is obtained that the maximal allowable $\delta$
tends to be infinity. It is worth noting that a very large $\delta$
generally implies a high-gain controller \dref{cl}.
For instance, the product of
the feedback gain matrix $K$ and the threshold $c_{th}$
corresponding to $\delta=1.5\times 10^5$
is obtained as $c_{th}K=-\begin{bmatrix} 1.2558 &0.0074 \end{bmatrix}\times 10^{5}$.

{\small \bf Example 2}. The dynamics of the discrete-time agents are given by \dref{di1}, with
$$A=\begin{bmatrix} 1 & 2 &0\\ 0 &1& 0\\ -1 &0 &-0.6 \end{bmatrix},B=\begin{bmatrix} 0 \\ 1\\1\end{bmatrix},
D=\begin{bmatrix} 0.8 \\ 0\\0 \end{bmatrix}, E=\begin{bmatrix} 0& 1 & 0\end{bmatrix}.
$$
The communication topology is given as in Fig. 1, with the first and last nodes knowing their own states.
In \dref{cldi},
let $\hat{d}_1=0.3$, $\hat{d}_6=0.5$, and $\hat{d}_i=0$, $i=2,\cdots,5$. Thus, the matrix $\widetilde{\mathcal {D}}$
in \dref{netdi} is
$$\widetilde{\mathcal {D}}=\begin{bmatrix} 0.1 & 0.15 & 0.15 & 0.15 & 0.15 &0\\
0.15 & 0.5 & 0.35 & 0& 0 &0\\ 0.15 &0.35 &0.3 &0 &0 &0.2\\
0.15 &0 &0 &0.1& 0.15& 0\\ 0.15 &0 &0& 0.15& 0.5 &0.2\\
 0 &0 &0.2 &0 &0.2 &0.1 \end{bmatrix},$$ whose eigenvalues
are -0.1611, -0.0644, 0.0959, 0.2257, 0.6316, 0.8722.
Using the Sedumi toolbox \cite{sturm1999using} to maximize $\delta$ in the LMI \dref{th5} with $\kappa=0.9$
yields the largest allowable certainty bound $\delta_{\max}=2.5$ and the corresponding solution are as follows:
$$\begin{aligned}
P &=\begin{bmatrix}    98.2213  & -2.0000 & -61.3883\\
   -2.0000  &  0.1197  &  1.3573\\
  -61.3883   & 1.3573  & 86.2810 \end{bmatrix},\\
W &=\begin{bmatrix}0  & -0.0780 &  -0.0612\end{bmatrix},
  \quad\tau=0.0912.
\end{aligned}$$
Thus, the feedback gain matrix of \dref{cldi}
is obtained as $K=\begin{bmatrix} -0.0195 &  -0.9888  &  0.0009\end{bmatrix}$.
The state trajectories of network \dref{netdi} are depicted in Fig. 3,
with the uncertainties $F_i$ randomly chosen within $[-2.5,2.5]$.

\begin{figure}[htbp]
\centering
\includegraphics[width=0.5\linewidth,height=0.3\linewidth]{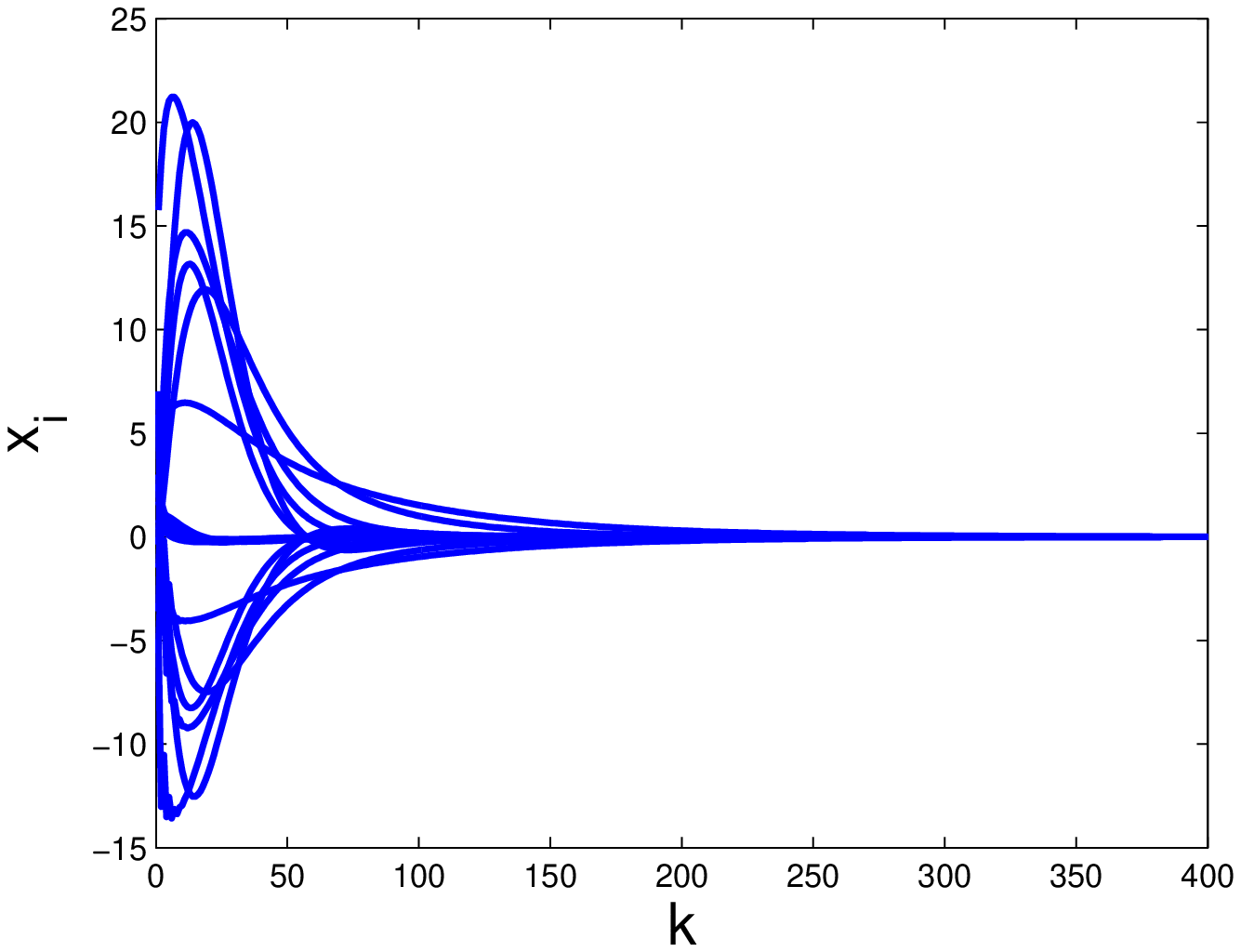}
\caption{The state trajectories of \dref{netdi} under controller \dref{cldi}. }
\end{figure}

\section{Conclusion}

In this paper, the distributed robust
control problems of uncertain linear multi-agent systems
have been considered, where
the agents are assumed to have identical nominal dynamics while subject to different
norm-bounded parameter uncertainties.
Distributed controllers have been designed,
based on the relative states of
neighboring agents and a subset of absolute states of the agents.
It has been shown for both the continuous- and discrete-time cases that the distributed quadratic stabilization
problems under
such controllers are equivalent to the $H_\infty$ control problems
of a set of decoupled linear systems having the same dimensions as a single agent.
Algorithms have been further presented to construct
the distributed controllers.
An important yet challenging topic for future research
is to extend the results of this paper to solve
the consensus and formation control problems of uncertain multi-agent systems.

\end{document}